\documentclass[runningheads]{llncs}

\usepackage{floatrow}
\newfloatcommand{capbtabbox}{table}[][\FBwidth]

\usepackage{todonotes}
\usepackage{fixltx2e}
\usepackage{amssymb}
\usepackage{graphicx}
\usepackage{wrapfig}
\usepackage{subfig}
\usepackage{subfloat}
\usepackage{url}
\usepackage{amsmath}
\usepackage{verbatim}
\usepackage{multirow}
\usepackage{paralist}
\usepackage{mathptmx}
\usepackage{times}
\usepackage{dirtree}
\usepackage{tabularx}
\usepackage{tikz}
\usetikzlibrary{arrows,%
                calc,%
                chains,%
                decorations.pathmorphing,%
                decorations.pathreplacing,%
                decorations.shapes,%
                graphs,%
                matrix,%
                positioning,%
                shapes.geometric,%
                shapes,%
                shapes.symbols,%
                trees
   			  }

\usepackage{graphicx}
\usepackage{pgfplots}
\usepackage{floatrow}

\usetikzlibrary{automata,arrows}
\usetikzlibrary{datavisualization}
\usetikzlibrary{datavisualization.formats.functions}
\usetikzlibrary{positioning, calc, backgrounds, arrows}
\usetikzlibrary{shapes.geometric}
\usetikzlibrary{decorations.pathreplacing}
\usetikzlibrary{patterns}

\let\olddefinition\definition
\def\definition{\olddefinition\footnotesize}

\newcommand{\optional}[1]{}

\usepackage[ruled,vlined,boxed,linesnumbered]{algorithm2e}
\SetKwRepeat{Do}{do}{while}
\SetKwInOut{Input}{input}
\SetKwInOut{Output}{output}

\graphicspath{{images/}}

\begin{document}

\title{Why Do Banks Find\\ Business Process Compliance So Challenging?\\ An Australian Case Study}
\author{	Nigel Adams\inst{1}
            Adriano Augusto\inst{1} \and
		    Michael Davern\inst{1} \and
			Marcello La Rosa\inst{1}
			 }
\titlerunning{ }
\authorrunning{ }

\institute{
           University of Melbourne, Australia\\
           naadam@student.unimelb.edu.au\\
   		    \{a.augusto, m.davern, m.larosa\}@unimelb.edu.au 
}

\maketitle 

\begin{abstract}
Banks play an intrinsic role in any modern economy, recycling capital from savers to borrowers. They are heavily regulated and there have been a significant number of well publicized compliance failings in recent years. This is despite Business Process Compliance (BPC) being both a well researched domain in academia and one where significant progress has been made. This study seeks to determine why Australian banks find BPC so challenging. We interviewed 22 senior managers from a range of functions within the four major Australian banks to identify the key challenges. Not every process in every bank is facing the same issues, but in processes where a bank is particularly challenged to meet its compliance requirements, the same themes emerge. The compliance requirement load they bear is excessive, dynamic and complex. Fulfilling these requirements relies on impenetrable spaghetti processes, and the case for sustainable change remains elusive, locking banks into a fail-fix cycle that increases the underlying complexity. This paper proposes a conceptual framework that identifies and aggregates the challenges, and a circuit-breaker approach as an ``off ramp" to the fail-fix cycle.

\end{abstract}

\section{Introduction}\label{sec:intro}

Banks play an intrinsic role in any modern economy, they recycle capital between savers and borrowers and are tightly regulated. In the last five years, Australian regulators have highlighted multiple compliance issues, particularly among the four major domestic banks, many of which are business process related. This has led to: i) a Royal Commission~\cite{Hayne2019Royal}; ii) regulators issuing penalties exceeding A\$2 billion\footnotemark; iii) tightening executive accountability\footnotemark[\value{footnote}]; \footnotetext{\label{note1}\scriptsize{\url{https://www.austrac.gov.au}, \url{https://www.apra.gov.au}}} iv) more than A\$8 billion in remediation costs and a significant investment in compliance resources\footnote{\scriptsize{\url{https://home.kpmg/au/en/home.html}, \url{https://www.robertwalters.com.au}}}; and v) the resignation of three CEOs and two Chairmen of these banks. 

It is not just Australian banks that struggle. Since 2008, US banks have been fined \$243bn for compliance-related events and the global cost of compliance for financial services firms is equivalent to an 8\% tax\footnote{\scriptsize{\url{https://www.ascentregtech.com/blog/the-not-so-hidden-costs-of-compliance/}}}. 

Academic interest in the field of business process compliance (BPC) traces its roots back to corporate scandals at organizations such as Enron, HIH, AIG, Lehmann Brothers and Société Générale along with the ensuing legislative changes (e.g., Dodd-Frank, Sarbanes-Oxley) at the turn of the millennium. The overarching challenge for BPC is to capture compliance requirements and check that business processes are operating in line with these requirements. This implies a need to evaluate processes throughout the BPC lifecycle: at design-time, run-time, and post-execution~\cite{Hashmi2018AreWeDoneYet}. While research efforts have focused on automating BPC, and significant progress has been made, there are still several research gaps~\cite{Hashmi2018AreWeDoneYet} and BPC is still highly manual and time consuming. 


In this setting, this paper investigates the reasons why banks struggle to keep up with BPC. We focus on the Australian banking context, and then discuss how the findings are generalizable to other banking contexts. 

To this end, first, we conducted a series of semi-structured interviews on BPC and its challenges. Next, following the Gioia methodology~\cite{gioia2013seeking}, we thoroughly analyzed the interview transcripts to identify factors inhibiting BPC. The interviews were conducted with participants drawn from the four major Australian banks, namely Australia and New Zealand Banking Group, Commonwealth Bank of Australia, National Australia Bank, and Westpac Group. The participants had backgrounds in \emph{Operations}, \emph{Risk \& Compliance}, \emph{Technology}, and \emph{Process Excellence}. 

In light of the above, this paper contributes a conceptual framework that identifies and aggregates 23 concepts capturing various challenges emerging from the interviews into seven key \textit{themes}, which are further grouped into three \textit{aggregate dimensions}. These dimensions are symptomatic of a fail-fix cycle that is entangling the banks. Based on this, the paper further suggests a circuit-breaker approach to address this cycle.

The remainder of this paper is structured as follows. Section 2 provides background to the study, including the regulatory context for Australia's major banks and the relevant BPC literature. Section 3 outlines our research methodology, results, and analysis. Section 4 discusses our findings while Section 5 presents the limitations of the study. Section 6 concludes the paper and discusses avenues for future work.
\section{Background and Related Work}\label{sec:background}
To address our research question, an understanding of both the BPC literature and the Australian banking context is required. Here, we provide a summary of both.

\subsection{Business Process Compliance} 
BPC is a well researched area. A BPC solution comprises multiple elements, each with a range of techniques proposed in the literature. At its core a BPC solution must demonstrate: an ability to capture requirements~\cite{deAraujo2013Automatic}; an approach and a language to formalize the rules~\cite{Governatori2009Journey,Elgammal2010Formal}; an approach to represent the process~\cite{Awad2008Efficient,Montali2014Monitoring}; a technique to check compliance between the process and the rules~\cite{Liu2007static,Doganata2009Effect} with regards to different process perspectives~\cite{Knuplesch2016Visually}. There are also a range of supporting features that enhance BPC's value such as: business reporting~\cite{Montali2014Monitoring}; violation handling~\cite{Maggi2011Monitoring}; feedback and root cause analysis~\cite{Ramezani2012Misbehave}; and change handling~\cite{Ly2012enabling}. Most contributions focus on one or more of these elements at a specific stage of the BPC lifecycle: \emph{design-time}; \emph{run-time}; and \emph{post-execution} (i.e., auditing). 


Framework-oriented solutions provide the backbone for BPC efforts and cover a broad spectrum, from enterprise-wide, high-level risk management frameworks (e.g., COSO) to industry and function specific frameworks (e.g., Basel accords) to frameworks that aim to solve a specific piece of the BPC puzzle, e.g., a taxonomy-based framework in~\cite{Rosemann2005Integrating}, or an evaluation framework in~\cite{Ly2013framework}. 

Managing BPC at design-time is a preventative strategy, concerned with ensuring that processes comply with relevant rules and regulations prior to execution -- either during the design-process~\cite{Governatori2009Journey} or post-design but pre-execution~\cite{Milosevic2006Towards}.
Debate has centred on approaches and languages that are expressive enough to handle the range and complexity of compliance requirements~\cite{Governatori2009Journey} but are seen to be technically complex, and those languages that are more business-user friendly (e.g., pattern-based approaches) but potentially lack some of the expressiveness~\cite{Elgammal2010Formal}. A range of techniques have been used to represent the process, such as Petri nets, UML diagrams, BPEL models, however, BPMN models are becoming the most popular in industry.


Run-time methods verify compliance during the process execution, and typically address aspects of BPC that cannot be verified and validated at design-time, e.g., ``segregation of duties'' and ``deadlines for completion'' requirements~\cite{Maggi2011Monitoring}. Proposed solutions fall into two broad categories: reactive, where compliance verifies progress-to-date~\cite{Maggi2011Monitoring}; and proactive monitoring, where progress-to-date knowledge is used to predict compliance outcomes~\cite{Leitner2010Monitoring}. While there are a range of techniques to capture the run-time process data, event streams are becoming the dominant approach~\cite{Leitner2010Monitoring,Maggi2011Monitoring}. However, this is challenging for processes producing sparse event streams, and computationally complex for processes producing large amounts of events in a short time.

Auditing is a post-execution strategy, traditionally both manual and based on sampling, there is now a shift to continuous auditing. Some of the auditing approaches covered in the BPC literature are based on process mining (PM) techniques~\cite{VanderAalst2010Auditing,Ramezani2012Misbehave,Doganata2009Effect}, these techniques benefit from reviewing a population of transactions instead of a sample. Database-driven solutions have also been proposed~\cite{Johnson2007Compliance,Agrawal2006Taming}. As a detective control, auditing does not prevent compliance breaches, but can be useful to inform process enhancements and also assess the impact of changed requirements. 

While much progress has been made in BPC research, there are outstanding challenges to apply the techniques in real world scenarios~\cite{Hashmi2018AreWeDoneYet,Becker2012Generalizability}. There is a recognition that the goal of automating BPC may be out of reach, with the focus now shifting to facilitation rather than full automation~\cite{Barnawi2016antipattern-based}. 

\subsection{Australian Banking Context}
There are 98 banks operating in Australia, controlling A\$5.2 trillion of assets\footnote{\scriptsize{\url{https://www.apra.gov.au}}}.
The four major banks account for 74\% of these assets. Over the last five years, multiple compliance breaches have been made public, highlighting a range of challenges the banks face in trying to maintain process compliance.

The findings of the Royal Commission and other regulators include: the extent of legislation banks are subject to; the difficulties banks have both understanding and interpreting them; blurred lines of accountability and bureaucratic decision-making; the extent of processing and administrative errors (A\$239m repaid in mortgages alone); poor processes; the age and complexity of product systems; a reactive approach to operational risk management and inability to detect systemic issues; a reliance on manual, detective controls that do not operate end-to-end; issues not addressed in a timely manner; trade-offs between funding compliance initiatives versus other initiatives; and an inability to ``join the dots"~\cite{APRA2018CBAPrudential,WBC2020Findings,ASIC2018Breach,Hayne2019Royal}.

\section{Methodology, Results, and Overview}\label{sec:app_results}
In this section, first, we introduce the Gioia methodology~\cite{gioia2013seeking} and  discuss how we applied it to the context of this study. Then, we report the results we obtained and provide a broad overview before discussing the results in depth in the next section.

\subsection{Methodology}\label{subsec:approach}
In this study we applied the Gioia methodology~\cite{gioia2013seeking}, given its ability to bring ``qualitative rigor" to the conduct and presentation of inductive and abductive research.  It provides guidelines to create a conceptual data structure comprising \textit{1\textsuperscript{st} Order Concepts} directly extracted from a set of interview transcripts, then analyzed and consolidated into \textit{2\textsuperscript{nd} Order Themes}, and finally distilled into \textit{Aggregate Dimensions}. Specifically, we executed the Gioia methodology by completing the following seven steps. 


\textbf{1. Develop Interview Protocol.} First we developed an \emph{interview protocol} to conduct the interviews. Development of the protocol was informed by a review of the BPC literature and enriched by an understanding of the banking context. It started with three introductory questions which sought participants views on what BPC meant, the impact of the recent regulatory issues, and which teams were impacted. The heart of the interview focused on: examples of the BPC issues the participants experienced, what they would do in hindsight, how they thought the issues could be prevented, how they would measure BPC performance, and the role they thought process mining could play to address the issues.

\begin{figure}[ht!]
  \includegraphics[scale=.56]{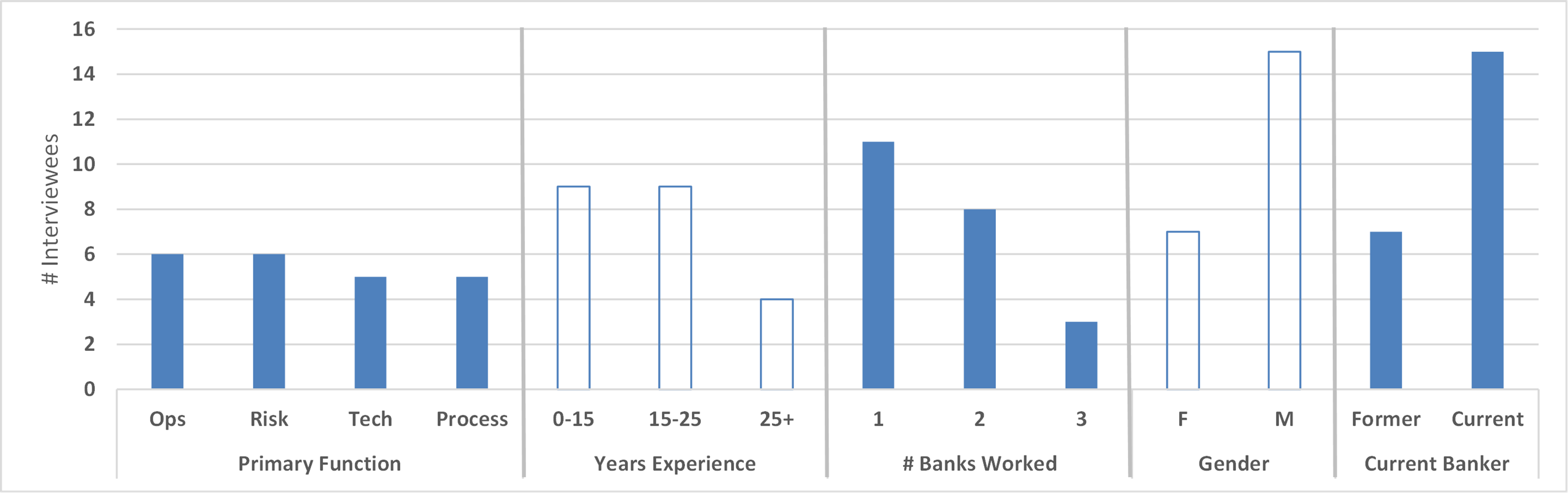}
  \caption{Interviewee Profile}
  \label{fig:interviewee_attributes}
\end{figure}

\textbf{2. Select Interviewees.} The industry participants in this study (i.e., the interviewees) were drawn from the authors' network. 54 potential interviewees were approached and 22, one-hour, semi-structured interviews were conducted. The interviewees were predominantly Senior Managers, Heads of, and General Managers within the relevant organization. All had banking experience in the last five years with at least one of the four major Australian banks. The average banking tenure (including international banking experience) was 17 years, and half of the participants had worked for more than one of the banks. Each of the banks was represented by at least five interviewees who had worked there. The interviewee profile is shown in Figure~\ref{fig:interviewee_attributes}.

\textbf{3. Conduct \& Transcribe Interviews.} We sent a \emph{Plain Language Statement} to participants prior to their interview, to provide them with context. All the interviews were conducted over Zoom calls, between June and December 2021. Initial pilot interviews were conducted with eight industry participants, to validate the line of questioning. Each recorded interview was subsequently de-identified and transcribed. 

\textbf{4. Code Transcripts.} The interview transcripts were imported into Nvivo and a word frequency count was run on the interviewees' responses to identify key terms based on both exact and stemmed word matches. The results were mixed. Some of the most frequently used terms had ambiguous meaning. For example, the word ``end" was frequently referenced regarding end-to-end process -- a salient term in this study, but it was also used as a figure of speech, e.g., ``at the end of the day", ``the end result". 

Hence, while the automated word frequency functionality provided some insight, its usefulness was limited to providing a base list of frequently used terms. This was addressed by manually coding the derived list of terms to each paragraph of the pilot interview transcripts, where they were used in the relevant context. The terms were also enhanced with participant-used synonyms. The result of this step created 152 codes, with some paragraphs associated with as many as 22 codes. 

The remaining 14 transcripts were subsequently coded. During this process an additional 17 terms were identified and added as new codes. No new codes were added for the last three transcripts suggesting saturation had been reached. Each transcript was coded to between 100 and 121 codes, with an average of 111 codes. The number of references per transcript ranged between 401 and 734 with an average of 511.

\textbf{5. Develop 1\textsuperscript{st} Order Concepts.} Following the initial coding, we then derived the 1\textsuperscript{st} Order Concepts through an iterative process. The 152 codes were consolidated by relying on the available data and the expertise of the authors of this paper. The first task was to consolidate terms into synonymous concepts. For example, terms such as ``audit", ``QA, QC" were consolidated with terms like ``controls" and ``checks". The results were reviewed by two co-authors, who provided suggestions for refinement, and then another iteration would start -- until no other changes were proposed. This iterative approach exposed two main concerns. First, some terms were applicable in a range of contexts, e.g., the most frequently used term was ``process'', but depending on the context it could refer to process design, a specific banking process such as mortgage lending, or process mining. Second, not all references to a term were indicative of a BPC challenge, e.g., some interviewees referred to a term in a favorable light. Taking these points into consideration, further iterations resulted in the terms being consolidated into 23 1\textsuperscript{st} Order Concepts.

\textbf{6. Validate Coding.}
To validate the coding, four transcripts were randomly chosen and coded separately to the 1\textsuperscript{st} Order Concepts by two authors. Only the interviewee's comments that referred to a concept as a challenge were coded. The results were compared for consistency, calculating the \emph{Cohen Kappa Coefficient} (CKC) for each transcript~\cite{cohen1968weighted}. The resulting CKC, for each of the four transcripts, was between 90\% and 93\%, suggesting that there was a high degree of consistency. 
The few mismatches mainly related to the authors' different expertise and background. A final round of validation was undertaken by the remaining two authors of the study. In this case, they reviewed selected text extracts for each of the 1\textsuperscript{st} Order Concepts.

\textbf{7. Develop 2\textsuperscript{nd} Order Themes and Aggregate Dimensions.}
The final stage to generate the output \emph{data structure} was to derive the 2\textsuperscript{nd} Order Themes and Aggregate Dimensions. To do so, we followed an iterative process that leveraged the industry experience of the first author\footnote{{\scriptsize{The first author is a former Senior Executive of two of the four banks studied, with more than 15 years of experience.}}} alongside existing theory and literature.


\subsection{Results}\label{subsec:results}
We describe the results in three layers: 1\textsuperscript{st} Order Concepts, 2\textsuperscript{nd} Order Themes, and Aggregate Dimensions. We begin with a description of the 23 1\textsuperscript{st} Order Concepts, and then discuss how we aggregated them into the 2\textsuperscript{nd} Order Themes and the Aggregate Dimensions. The data structure summarizing our results is reported in Table~\ref{fig:data_structure}.

\begin{table}[htbp]
\def\arraystretch{1.2} 
\setlength{\tabcolsep}{4.5pt} 
{\scriptsize{
  \centering
  \caption{Resulting \emph{Data Structure} from our analysis. The last two columns respectively report the references per concept and the number of interviewees referencing a concept.}  \label{fig:data_structure}%
    \begin{tabular}{lcccc|cc}
    \hline
    
    \multirow{2}{*}{\textbf{1\textsuperscript{st} Order Concepts}}
    & \multirow{2}{*}{$\rightarrow$}
    & \multirow{2}{*}{\textbf{2\textsuperscript{nd} Order Themes}} 
    & \multirow{2}{*}{$\Rightarrow$}
    & \textbf{Aggregate} 
    & \multicolumn{2}{c}{\textbf{\#}}\\

    & 
    & 
    & 
    & \textbf{Dimensions} 
    & \textbf{Ref.}
    & \textbf{Int.}\\
    
    \hline
    Complex Model & $\rightarrow$ & \multirow{4}{*}{\shortstack{Dynamic \& \\Complex Ecosystem}} & \multirow{7}{*}{$\Rightarrow$} & \multirow{7}{*}{\shortstack{Complex \& \\Dynamic \\Requirements\\Load}} & 72& 21\\
    Regulatory Pressure Intensifying & $\rightarrow$ &       &  & & 84& 18\\
    Disruptive Competition & $\rightarrow$ &       &  & & 19& 10\\
    Frequently Changing Direction & $\rightarrow$ &       & &  & 22& 9\\\cline{1-3}    
    
    Multiple Requirement Types & $\rightarrow$ & \multirow{3}{*}{\shortstack{Complex \\Requirements}} &  & & 30& 20\\
    Translating Ambiguous Requirements & $\rightarrow$ &       &  & & 37 & 14\\
    Conflicting Objectives & $\rightarrow$ &       &  & & 64& 19\\\hline
    
    Inflexible, Disconnected Legacy Technology & $\rightarrow$ & 
    \multirow{3}{*}{\shortstack{Disjointed \& Disparate\\Process Foundations}} & \multirow{10}{*}{$\Rightarrow$} & \multirow{10}{*}{\shortstack{Impenetrable \\Spaghetti \\Processes}} & 71& 21\\
    Data Oasis, Information Mirage & $\rightarrow$ &       & & & 66& 17\\
    Fragmented Processes & $\rightarrow$ &       &  & & 72& 19\\\cline{1-3}
    
    Inadequate \& Ineffective Support & $\rightarrow$ & \multirow{3}{*}{\shortstack{Hard to Follow\\ Processes}} & & & 61& 16\\
    Too Many Exceptions & $\rightarrow$ &       & & & 38& 17\\
    ``Band-Aids", Patches \& Workarounds & $\rightarrow$ &       &  & & 51& 19\\\cline{1-3}    

    Huge Scale & $\rightarrow$ & \multirow{4}{*}{\shortstack{Resource Intensive\\Processes, Prone to Fail}} &  & & 18& 12\\
    Partial Automation Relies on People & $\rightarrow$ &       &  & & 72& 21\\
    Layers of Flawed Controls & $\rightarrow$ &       &  & & 106& 21\\
    Lack of Knowledge \& Experience & $\rightarrow$ &       &  & & 117& 20\\\hline
    
    System Monitored Not Processes & $\rightarrow$ & \multirow{4}{*}{\shortstack{Decision-Making \\Blind Spots}} & \multirow{6}{*}{$\Rightarrow$} & \multirow{6}{*}{\shortstack{Elusive Case \\for Sustainable \\Change}} & 29& 13\\
    Impaired Line of Sight & $\rightarrow$ &       & &  & 92& 20\\
    Change Execution Credibility & $\rightarrow$ &   &    &  & 82& 20\\
    Short-Sighted Investment & $\rightarrow$ &       & &  & 43& 14\\\cline{1-3}    

    Unclear Accountability & $\rightarrow$ & \multirow{2}{*}{Cultural Headwinds} &  & & 86& 20\\
    Tick-the-Box Culture & $\rightarrow$ &       &  & & 78& 20\\\hline
    \end{tabular}%
  }}
\end{table}%

\textbf{1) Complex model.} The banking business model is complex. There are a large number of products and services offered through multiple channels to a wide-range of customer segments across multiple jurisdictions, which are organized around multiple business units as part of an ecosystem dependent on many and varied 3rd party stakeholders. It is not just the dimensions of the model but the interconnected web that they form, e.g., products and processes that cut across organizations, business units and segments, and that are largely invisible and intangible. This complexity translates into a significant number of requirements that must be captured in process design.

\textbf{2) Regulatory pressure intensifying.} Interviewees focused on the changing nature of regulatory relationships, larger fines, more requirements and increasing scrutiny, or as Interviewee-1 put it: ``[...] it's just a wave after wave of regulation". Interviewees also referenced: the cost of compliance impacting competition; the regulators' product knowledge limitations; the fact that the burden of compliance is being felt directly by customers; the expanding role banks are expected to play helping police financial crime; and the perception that the regulatory relationship with the banks is adversarial, whereas a more collaborative approach is required to address many of the industry's issues.
    
\textbf{3) Disruptive competition.} Changing industry dynamics are also creating more requirements. New players are not burdened by inflexible technology, are far more agile -- introducing new products and features at a faster rate -- and in some cases are more lightly regulated. This is seen to present a cost advantage to non-bank participants but an increase in risk to the overall system.
    
\textbf{4) Frequently changing leadership and direction.} Staffing, structure and strategy changes, particularly at senior leadership levels, create work. It is not restricted to major changes but also more routine business decisions such as individual leaders changing roles, reducing project budgets or changing their risk appetite. These lead to a change in objectives, projects being re-scoped in-flight and resources re-distributed.

\textbf{5) Multiple requirement types.} In terms of the types of requirements that must be fulfilled, the initial response, for almost all participants, was to focus on regulatory obligations and requirements. However, follow-up questions revealed that there are many other types of requirements: industry codes of practice such as SWIFT standards; business policy; contractual obligations; and, of course, customer requirements. Variation also affects the necessity of a requirement (e.g., mandatory or ``nice to have"), as well as the consequences of failure (e.g., a significant fine or an adverse performance indicator). 

\textbf{6) Translating ambiguous requirements.} The way requirements are communicated is not always clear. They can be contradictory, duplicated, written in ambiguous language subject to both interpretation and translation, e.g., ``We need to do the right thing". In some cases the requirements are not known or not communicated. 

\textbf{7) Conflicting objectives.} Interviewees referenced the focus on sales, responsiveness, SLAs, cost efficiencies, and meeting the needs of investors first as higher priorities than quality or compliance. Even the threat of larger fines is perceived to be insufficient to change the mindset. There is also conflict between different teams in a bank highlighted by Interviewee-2: ``[...] we’re supposed to be innovative [...] but the brakes are put on by the compliance guys". 

\textbf{8) Inflexible, disconnected, legacy technology.} The technology environment comprises a multitude of in-house built systems and others sourced from multiple vendors. They are heavily customized, do not integrate easily, and are hard, slow and expensive to change, or as Interviewee-18 put it: ``We are sort of bound by the legacy system. To make a simple change in the system, it is quite difficult." Whereas Interviewee-5 stated: ``[As a roadblock] the one that springs to mind straightaway is integration. Our technology architecture is far from simple. There's bits and pieces logged [sic] all over the place." Interviewees also referenced individual processes dependent on 30 applications and a technology landscape with hundreds of disconnected systems.

\textbf{9) Data oasis, information mirage.} While there is no shortage of data, classifying it and accessing it, particularly at the right level of granularity, is not easy and there are also integrity issues such as: duplication, blind-spots, missing data, or data getting lost during migrations. Without unique identifiers and standards such as naming conventions, data does not flow easily between technology assets and processes, and stitching it together is expensive and time consuming.  
    
\textbf{10) Fragmented processes.} Processes are seen as something that occur within a function, not end-to-end. As the organization changes, the process boundaries also change. Over many years this has led to significant fragmentation with bits of processes dispersed across many teams and limited understanding of how the component parts fit together. Interviewee-8 stated: ``[...] we're constantly breaking up our processes to fit with a design that’s based around where people work, not what they do".

\textbf{11. Inadequate \& ineffective support.} "They've got really complicated checklists that they just don't use." (Interviewee-6) is one reference to the inadequacy of the tools and documentation process participants are working with. Others include: process models developed by people with limited process modeling skills, documentation that does not exist, is not maintained or only covers the ``happy path". In other cases, process maps are documented for the regulators, not for the teams operating the processes, or as an overlay, not an integral part of the process. Interviewee-11 commented: ``I know banks spend a heap of time and heap of budget on documenting processes. But if you ask the average person on the ground, they would say they're not documented."
    
\textbf{12) Too many exceptions.} Referring to a lending process, Interviewee-14 commented: ``You thought you had a 70\% STP [straight-through-processing rate]. The reality is you had 3\% because the other 97\% were taking one of the 56,000 variable pathways". The exceptions are typically driven by customizing siloed applications and insufficient project funding or a need to meet a customer request quickly. Exceptions are also introduced based on local considerations, e.g., the degree of latency or staff trying to navigate an easier path through the process. 

\textbf{13) ``Band-Aids", patches \& workarounds.} Years of under investment and constant tweaks, tinkering and point solutions have left processes strewn with workarounds, patches, bottlenecks and hand-offs. With a ``don't fix 'til it's broken" mindset (Interviewee-10), the ensuing urgency leads to more patches and workarounds. It also contributes to high rework rates and errors as the processes do not flow smoothly. Inadequate workflow was raised by multiple interviewees. Some refer to a workflow based on email and collaboration tools, others comment that only part of the end-to-end process has been workflow-enabled, negating the benefit.

\textbf{14) Huge scale.} While the scale varies by process, the high volume of time-critical transactions is frequently referenced, e.g. payments. Seasonality effects are also prevalent. It is not just high volumes; interviewees referenced the number of staff -- between 30,000-50,000 per bank, instances of putting hundreds of risk controls, and concurrent onboarding of 90+ new employees into a single team.


\textbf{15) Partial automation relies on people.} High-volume processes are only partially automated, resulting in a significant number of manual, repetitive tasks. It is not just the higher likelihood of errors, but the ability to absorb the degree of change and the lack of audit trails/visibility that prove challenging. Interviewees see automation as a panacea. 

\textbf{16) Layers of flawed controls.} The second most referenced concept, adding more controls is seen as the response to any problem. Many controls are after-the-fact and manual, many are flawed (e.g., 4-eye checks), many rely on sampling. Some are so complicated that they are not applied and the layering of controls mean that many are never activated. The lack of preventative controls is attributed to the fact they would slow the process down by multiple interviewees. 

\textbf{17) Lack of knowledge \& experience.} The most frequently referenced 1\textsuperscript{st} Order Concept is driven by: a loss of knowledge and experience as long-tenured staff leave; the narrowing of focus to learn discrete tasks instead of the end-to-end process; the difficulty in attracting talent to work with legacy technology; the time it takes to train new people (up to twelve months); and the workload pressure the teams face. 

\textbf{18) Systems monitored, not processes.} Business activity monitoring and systems monitoring are referenced by multiple interviewees, but while they may trigger alarms in terms of queue depth and system performance they provide no indication of process performance. Interviewee-3 commented: ``So most of our [operational technology] monitoring has a focus on system health and availability [...] in terms of monitoring actual processes, we don't really have that in place." 

\textbf{19) Impaired line-of-sight.}  Metrics and reporting do not link business outcomes to process performance and systems events. Different teams look at different metrics, with different objectives, hence decision-making is challenging. Some interviewees believe the data is there, but it has never been considered important enough to extract. Some assert the data exists but does not translate into decision-making information. It is not just about availability, but also about actionability for real-time decision-making. 

\textbf{20) Change execution credibility.} With processes so dispersed and so many stakeholders, getting buy-in, managing the various self-interest groups and capturing requirements up front is challenging. Communicating the change and the implications are also seen by interviewees as gaps. Interviewees referenced projects running over budget and then being re-scoped, typically leading to more workarounds. More worryingly were ``improvement" projects that made things more complicated, unwittingly removed key controls, did not deliver or did not actually fix the problem. As Interviewee-19 noted: ``I've been here 11 years and I have not known one [of 20 projects] fix the problem statement that we need to fix." Given the change process can be slow, unofficial, shadow change processes also play a role -- which leads to more tinkering.

\textbf{21) Short-sighted investment.}  The reason for not progressing the business case is typically referenced as: the investment is too high, the time to make the changes too long, the benefits and potential value are unclear and too far in the future when results are needed now. In other words, the sustainable, strategic business case does not stack up relative to the alternative options of more FTEs\footnote{Full Time Equivalent resources}, patches and remediation. Interviewee-11 added an additional insight: ``Senior leaders [...] want to work on the strategic stuff. I'm not sure processes are seen as strategic enough."

\textbf{22) Unclear accountability.} Accountability is confused, particularly with regards to the three lines of defense model. One incident referenced involved an account owner, an ATM network owner, a branch owner, a product owner, and a customer owner. There was no reference to a process owner. Progress has been made, while the BEAR legislation does not attribute process ownership directly, it at least makes it implicit, and there has been investment in bolstering risk and compliance support resources. However, interviewees believe that this has led to accountability being removed one step further from the source of risk. The propensity to engage more consultants and lawyers has had the same effect. 

\textbf{23) Tick-the-box culture} When asked who is responsible for compliance, ``it's everyone's job" was a common response, with the heightened scrutiny creating a sense of nervousness. However, the message becomes confused and appears to lose momentum as it filters down through the organizational hierarchy.  Throughout the transcripts there are references to BPC being perceived as a toll-gate, a "tick-the-box" exercise, people mechanically following a process whether it is right or wrong to avoid blame -- even though consequence management is rare -- people not challenging the status quo or afraid to speak up and BPC tasks executed with a sense of complacency. It is not seen as strategic but it is important to be seen to act.

Synthesizing these 23 \emph{1\textsuperscript{st} Order Concepts} together led to the identification of seven \emph{2\textsuperscript{nd} Order Themes}. The complex business model, intensifying regulatory pressure, disruptive competitive landscape and the impact of leadership and directional changes combine to highlight a \emph{Complex \& Dynamic Ecosystem} generating a significant number of \emph{Complex Requirements} -- there are many different types, they are frequently ambiguous, and they must cope with conflicting objectives and priorities. This requirement load is imposed on \emph{Disjointed \& Disparate Process Foundations}, where the underlying legacy technology is poorly integrated and inflexible, the data is plentiful, but hard to access and limited in its ability to convey information, and processes are repeatedly re-aligned to follow organizational structure changes. Additionally, the lack of support materials and the extent of exceptions and repeated patching lead to \emph{Hard to Follow Processes}. Despite such challenging foundations, the volume going through the processes is significant, yet they are only partially automated and not error-proofed, hence they are \emph{Resource Intensive Processes, Prone to Fail}, exacerbated by the fact that there is a lack of knowledge and experience. Not monitoring end-to-end processes thwarts the ability to align process performance and business objectives, and with a poor track-record executing change these \emph{Decision-Making Blind Spots} hinder investment. There are also \emph{Cultural Headwinds}, where ownership and accountability for addressing the issues is unclear and BPC is seen as ``tick-the-box" exercise instead of a strategic endeavor.

These seven \emph{2\textsuperscript{nd} Order Themes} were further consolidated into three \emph{Aggregate Dimensions}. A complex and dynamic ecosystem coupled with a high degree of requirements complexity leads to a \emph{Complex \& Dynamic Requirements Load}. This demand on the organization is fulfilled by \emph{Impenetrable Spaghetti Processes}, where resource intensive processes are hard to follow, prone to fail and built on disjointed and disparate process foundations. Decision-making blind spots across end-to-end processes and cultural headwinds make for an \emph{Elusive Case for Sustainable Change}. Together, these dimensions provide insight into why the major banks find BPC challenging. 


\subsection{Overview}\label{subsec:analysis}
The purpose of this study is to understand why banks find BPC challenging. It should be noted that bank processes are not homogeneous and interviewees also provided examples where, in specific cases, some of the \emph{1\textsuperscript{st} Order Concepts} were being, or had been addressed. However, for those processes experiencing BPC challenges, the results above represent the common themes.

From this point on, we refer to individual \emph{1\textsuperscript{st} Order Concepts} with a ``C" and their ID number as listed above.
Overall, 48\% of references were associated with the \emph{Impenetrable Process Spaghetti} aggregate dimension, 29\% with \emph{Elusive Case for Sustainable Change}, and 23\% with \emph{Complex \& Dynamic Requirements}. The number of references per \emph{1\textsuperscript{st} Order Concept} ranged from 18 to 117 (see also Table~\ref{fig:data_structure}), with the top five concepts accounting for 34\% of the references (C17, C18, C19, C22, C2 respectively) and the top eleven accounting for 66\%. 13 \emph{1\textsuperscript{st} Order Concepts} were referenced by more than 18 of the interviewees, and only two concepts were referenced by less than half of the interviewees (C4 and C3 respectively). 

In terms of coding differences, with a limited number of exceptions, interviewees' responses were relatively homogeneous. There were no material differences (within $\pm$5\% of the average) by gender and current role. Those with more than 25 years of tenure focused less on \emph{Impenetrable Spaghetti Processes} while those who had worked in three of the banks placed more emphasis on \emph{Complex \& Dynamic Requirements Load} (+8\%) and less on \emph{Elusive Case for Sustainable Change} (-9\%). By function, Operations and Process Excellence interviewees focused more on \emph{Impenetrable Spaghetti Processes} (+8\% and +6\% respectively). The Risk \& Compliance interviewees emphasized \emph{Complex \& Dynamic Requirements Load} (+10\%) and Technology interviewees favored \emph{Elusive Case for Sustainable Change} (+7\%). This was offset by Risk \& Compliance and Technology interviewees placing less emphasis on \emph{Impenetrable Spaghetti Processes} (-6\% equally) and the Process Excellence interviewees focusing less on \emph{Complex \& Dynamic Requirements Load} (-6\%).

The implications of these results are discussed in the next section.

\section{Discussion}\label{sec:discussion}
Some of the challenges associated with the 1\textsuperscript{st} Order Concepts may sound familiar. The BPC literature refers to: the difficulty in extracting and translating ambiguous requirements (C5, C6); the importance of process ownership and clear roles and responsibilities (C22); aligned metrics and reporting (C19); managing scale (C14); incorporating real-time monitoring (C18); the importance of managing change well (C20); the importance of proactive, preventative controls and compliance-driven design (C16); and the relative merits of annotating (overlaying) versus integrating controls in process models (C11). The business process management (BPM) community is more than familiar with the impact of variation (C12); the importance of good documentation (C11); adequate training (C11); fit for purpose automation (C15); legacy integration (C8); and data-connectivity issues (C9).

There are many solutions available to (partially) address these issues, and it is tempting to suggest appointing a team of process professionals to at least fix the process-related issues. However, as Interviewee-8 pointed out: ``Every three to five years the banks get rid of all their process improvement people." Why is this the case?

Interviewees referenced the invisible, intangible nature of processes in banking, making it harder to see when and where a process has failed. Others referenced the fact that the banks do not control the end-to-end process (e.g., in payments and broker originated mortgages), hence, managing input quality with ecosystem partners is harder. However, there are examples of solutions cited in the literature referencing common bank processes --  the same processes discussed by the interviewees: account opening, lending, payments, and customer onboarding.
Another interviewee referenced the fact that there are risk people and operations people but no risk operations people. Again, the BPC literature assumes that these are people with separate skill sets working collaboratively, as has been the case in the major banks, so there should be no impediment.

The answer may lie in four of the most referenced terms: ``end-to-end", ``complexity", ``perspective", and ``understand" with 154, 141, 205 and 270 references respectively. The terms appear in multiple contexts, which have been assigned to the most relevant \emph{1\textsuperscript{st} Order Concept}, but the overarching theme is that, from the interviewees perspective, end-to-end processes are so complex that people do not understand them. Interviewee-1 summed it up by saying: ``So I think this spaghetti, this complexity that underpins very aged infrastructure with Band-Aids plastered across it. The lack of knowledge. Who understands that [...] when it falls over? And people apply another fix and another fix." Interviewee-11 elaborates on this: ``I think complexity comes from [...] the fact that no one knows what's going on", while Interviewee-8 states: ``Most people [...] do not understand the fundamentals of what a process is" and goes on to refer to process myopia. 

This is supported by references to the inadequacy of the tools and techniques used to help people understand their processes in the process identification, discovery, and analysis phases of the BPM lifecycle. Examples include: process maps that do not reflect reality, a lack of modeling skills, a reliance on subject matter experts who do not understand their processes nor their requirements, time (and expertise) constraints that mean the exceptions are not mapped, and the lack of performance data to inform the analysis of the current state. Our observation is that it is hard to see and analyze the system holistically, and people resort to solving in their own silos.

Moreover, when tasked with improvement, interviewees referenced three pathways: i) the complexity is underestimated in the project and the cost and timelines blow out, so the project is shut down; ii) the complexity is acknowledged in the project, cost is incurred but results are not delivered in a timely manner, so the project is shut down; and iii) an improvement to part of a process variant that is so immaterial no one notices. 

\begin{figure}[ht!]
\centering
\includegraphics[width=1.05\textwidth]{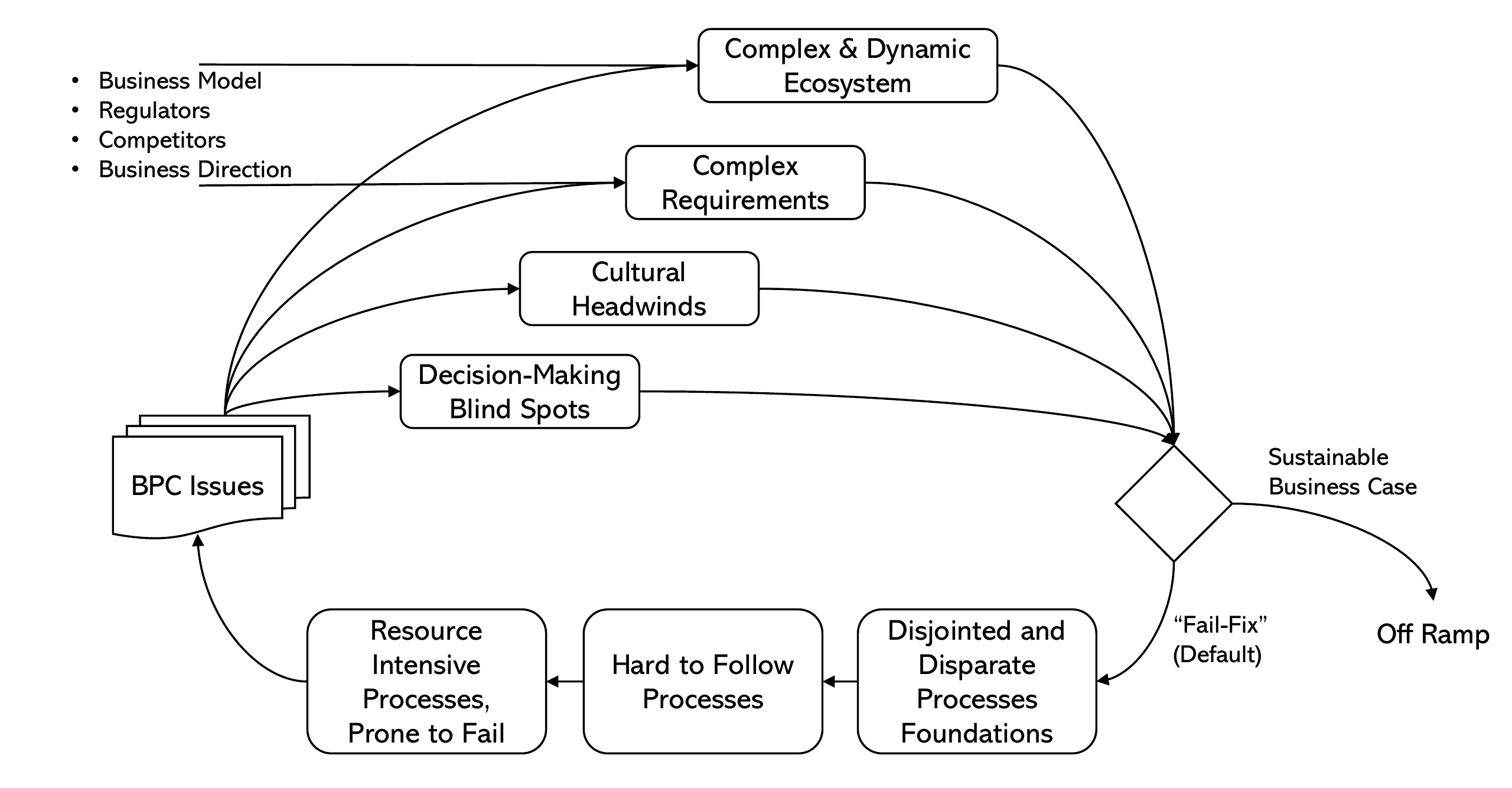}
  \caption{BPC Fail-Fix Cycle}
  \label{fig:cycle}
\end{figure}

Investment in a sustainable solution appears to be elusive. This can be attributed to: limited progress-to-date, ``siloed thinking", an inability to ``see" the intrinsic, system-wide costs of the spaghetti processes, the difficulty in ``selling" this type of business case, the level of investment required to fix it and a nebulous benefit case. Hence, the fail-fix approach persists, considering also the excessive requirements' load, the complexity increases. This fail-fix cycle emerged from the analysis of the interviewee transcripts. Our conclusion is that, while there may be solutions to the \emph{1\textsuperscript{st} Order Concepts} individually, an effective solution requires a more holistic perspective, and hence a shift to seeing those first order challenges through the lens of the \emph{2\textsuperscript{nd} Order Themes}. We elaborate this perspective in Figure \ref{fig:cycle}.

Our analysis of insights from the interviewees points to three fundamental challenges to break the cycle: i) end-to-end visibility; ii) effective collaboration across the ecosystem to simplify requirements; and iii) developing a viable, sustainable business case. In the following, we comment on each of these circuit breakers.

\textbf{End-to-end visibility.} 
The first circuit breaker is to focus on end-to-end visibility. As Interviewee-8 put it: ``If you can't see into your process, [...] you're running blind." Banks must be able to see the complexity to chart a course to untangle the spaghetti. However, this is not a traditional process discovery and analysis approach. The limitations discussed above preclude the traditional approach in favor of \emph{automated process discovery}~\cite{dumas2018fundamentals}, a core process mining capability. Seeing all the exceptions, all the pathways, all the performance data associated with those pathways is essential to improve transparency, enable visibility and enable analysts to view the system holistically.  Most interviewees were enthusiastic about the approach, and in some cases pilot process mining projects were already underway. 

The roadblock, as many interviewees pointed out, is accessing the event data across end-to-end processes and addressing the data blind-spots when partially automated processes reverted to manual tasks. Extract, transform and load (ETL) techniques and tools have certainly advanced in recent years, but stitching together data without a common identifier, with blended coarse- and fine-grain events, will require further work. 

Assuming reliable end-to-end event logs can be generated, a criticism of the automated process discovery approach is that it tends to produce spaghetti models -- 56,000 paths through a lending process is a case in point~(Interviewee-14). While it is essential to be able to see the full complexity, it is critical that techniques and tools also simplify abstraction. Interviewees refer frequently to \emph{layers} and \emph{perspectives}. The Risk \& Compliance function want to see the process through the controls lens or the regulatory lens, operational teams want to see the resource impacting flows and queues/bottlenecks. Our observation is that automated process discovery tools should provide a single model, capable of capturing the different user perspectives.

Interviewees reflected that leveraging event logs presents a range of other process mining opportunities, particularly, online conformance checking and variant analysis would help address C16 and C12, respectively, while automated process discovery also enables the other circuit breakers. 

\textbf{Effective collaboration across the ecosystem to simplify requirements.}
The second circuit breaker is to simplify requirements. At present, requirements are treated in silos, (e.g., regulatory requirements, codes of practice requirements, business policy and customer requirements). However, there is a significant amount of overlap. A small number of control patterns~\cite{Elgammal2010Formal} can be implemented to cover the majority of requirements. Because there are so many of them that are added incrementally, it is difficult to see through the control clutter and understand which ones are triggered in which circumstances and sequence. As Interviewee-15 put it: ``You'll have multiple risk controls that are never activated, therefore they're useless" and they go on to say ``I think the more controls you put in place, the higher the risk of any process, because you can add complexity". End-to-end visibility as described above will help users see the requirements complexity as well as understand its impact. It will also highlight the opportunity to standardize requirements registers, and risk \& control libraries. This then makes it easier to consolidate and prioritize rules and controls across the portfolio of requirements.

There is also an opportunity to work with ecosystem participants including regulators, industry bodies, and competitors to support simplification through standardization. Throughout the interviewee transcripts there are references to a lack of standards such as naming conventions and acronyms and terminology applied inconsistently. As well as developing a more rigorous suite of standards, taxonomies and ontologies, it is also critical to ensure that they are applied. An interviewee gave an example where member banks could choose to apply a code (which would enable straight-through-processing) but many chose to enter free text (which would generate an exception).  

\textbf{Developing a viable, sustainable business case.}
End-to-end process visibility also enables the third circuit breaker -- developing a viable business case for sustainable change. The fail-fix cycle is perpetuated not because of a lack of funding and resources, but because the business case to address the issues sustainably does not \textit{appear} to be viable. Interviewees refer to the level of investment required, the fact that senior leaders do not perceive BPC to be strategic, the lack of process ownership (sponsorship), that product and process costs are not known and the value of doing this well and sustainably are unclear. Leveraging the enhanced end-to-end visibility, the task is to demonstrate the value of a strategic approach to BPC: the impact of exceptions on the level of operating cost and operational risk capital held, the cost of re-hiring and training people on broken processes, the opportunity cost of not being able to absorb more change, the ``tech debt" associated with hollowing out legacy systems and keeping them on life support, the impact on product pricing of actually knowing your costs. It is hard to answer these questions accurately in the major banks today.

The second part to this circuit breaker is to package up programs of work that can deliver value in the financial periods the banks are beholden to. We asked interviewees where would they start if they were in charge of addressing the issues. The majority focused on piggy-backing off existing programs of work. They would select a process that was material to the bank's results, a significant pain-point for executives, one that is fully funded and with a senior executive sponsor already in place. They were adamant it should not be a program in its own right. The objective should be to enable an existing program to deliver better outcomes faster by addressing the issues discussed.

\section{Limitations of the Study}
There are several potential limitations of this study: we only interviewed 22 current and former staff members of Australia's four major banks. These banks collectively employ over 150,000 FTEs. However, the interviewees  represent key categories of process compliance stakeholders identified in the literature, namely Operations, Risk \& Compliance, Technology and Process Excellence. It is also important to note that saturation was reached after 19 interviews. 
In selecting our sample we wanted to ensure that interviewees were close enough to the point of execution to produce detailed anecdotes, but not so close that they could not see the broader context. To this end, the study focused on senior management. The executive layers, we determined, were too far from the detail and the process operators' focus was too narrow. Extending the study to cover other perspectives within the hierarchy is a potential future work stream.

This case study focused on four Australian banks. There is a risk that the findings cannot be inferred for other banks either within Australia or in other jurisdictions or other financial services participants. However, there is evidence that both the problems and the challenges apply to these other entities and in other geographies. For example, the Royal Commission found similar issues across the broader Australian financial services industry. As previously noted, other countries have also experienced similar situations. However, more work is required to validate this assumption. While significant effort has been undertaken to validate the results as described in the approach, there is still a degree of subjectivity in interpreting the findings, where the experience of the first author was relied on.
\section{Conclusion}\label{sec:conclusion}
Through a series of semi-structured industry interviews, our study has identified 23 \emph{1\textsuperscript{st} Order Concepts}, linked to seven \emph{2\textsuperscript{nd} Order Themes} and three \emph{Aggregate Dimensions} that help explain why major Australian banks find business process compliance so challenging. Challenges associated with some of the \emph{1\textsuperscript{st} Order Concepts} are known and well researched. The balance tend to be seen as organizational issues beyond the scope of this type of study. The most frequently mentioned concerns refer to i) a lack of knowledge and understanding of people involved in trying to establish and maintain business process compliance, followed by ii) a lack of preventative controls, and iii) a lack of visibility across end-to-end processes. However, our findings suggest that treating each of these concerns as a stand-alone issue will not address the underlying problem -- the major banks have already tried this approach. 

What makes our study different is the focus on addressing the overarching issue of the \textit{intrinsic complexity}. It is a term referenced throughout the interviewee transcripts in a wide range of contexts. The current approach of the banks, particularly how they approach process discovery and analysis, locks them into a negative fail-fix cycle of increasing complexity, which demands a circuit-breaker. We propose three areas to focus on to break the cycle: i) enhancing end-to-end visibility through automated process discovery; ii) simplifying requirements across the ecosystem; iii) and developing a viable, sustainable business case. 

It is important to note that the results of our study do not imply that every process, in each of the banks studied, exhibits all of these concerns. Counterexamples were also provided by interviewees, and each \emph{1\textsuperscript{st} Order Concept} should be seen as one end of a spectrum. Future work will focus on determining the spectrum range for each concept and develop a process profiling approach to help banks determine the extent of the business process compliance challenge by process. Further work is also required to determine the role of specific process mining techniques such as conformance checking and variant analysis and how they can address the challenges identified in this study. 

\smallskip\noindent\textbf{Acknowledgments:} We thank the interviewees who generously gave their time and shared their thoughts to help produce this paper.


\bibliographystyle{plain}
\bibliography{QUT,Banking}

\end{document}